\documentclass[prl,twocolumn, superscriptaddress]{revtex4}
\usepackage{graphicx,amsmath,amssymb,amsxtra}
\renewcommand{\narrowtext}{\begin{multicols}{2} \global\columnwidth20.5pc}

\def\be{\begin{eqnarray}}
\def\ee{\end{eqnarray}}

\newcommand{\Eq}[1]{Eq.~(\ref{#1})}

\begin{document}

\title{Halperin $(m, m',n)$ bilayer quantum Hall states on thin cylinders}


\author{Alexander Seidel}
\affiliation{National High Magnetic Field Laboratory, Florida State
University, Tallahassee, Fl 32306, USA}
\affiliation{Washington University, St. Louis, MO 63136, USA}
\author{Kun Yang}
\affiliation{National High Magnetic Field Laboratory, Florida State
University, Tallahassee, Fl 32306, USA}

\date{\today}

\begin{abstract}
The Halperin $(m,m',n)$ bilayer quantum Hall states are
studied on thin cylinders. In this limit, charge density
wave patterns emerge that are characteristic of the underlying
quantum Hall state. The general patterns are
worked out from a variant of the plasma analogy. Torus degeneracies are recovered,
and for some important special cases a connection to well-known
spin chain physics is made. By including interlayer tunneling, we also work out the critical
behavior 
of a possible phase transition between the $(331)$ state and the
non-abelian Moore-Read state in the thin cylinder limit.
\end{abstract}

\maketitle


A very successful strategy to solve problems in condensed matter
physics is to identify the correct simple state that
a given interacting quantum system can be smoothly evolved into.
On the other hand, our understanding of fractional quantum Hall states
is based on radically new techniques\cite{laughlin}.
Recently, however, a particular
way has been discussed to adiabatically transform fractional
quantum Hall liquids into trivial charge density wave (CDW)
states.
In Refs. \onlinecite{seidel1, seidel2, karlhede1, karlhede2}
it was observed that for certain Hamiltonians, a quantum Hall
ground state does not undergo a phase transition when the two-dimensional
surface of the system is deformed into a quasi one-dimensional (1D)
limit, e.g. a thin torus or cylinder.
In this limit,
simple CDW patterns
emerge\cite{haldane1}.
These patterns
are characteristic of the underlying quantum Hall state,
and can serve as ``labels'' for the
ground states and their elementary excitations.
Despite their seeming simplicity,
these labels turn out to be very efficient bookkeeping
tools.
They give rise to simple and picturesque
explanations for some abstract concepts in quantum Hall
physics, such as the existence of fractional charges and
topological sectors.
Braiding statistics can also be derived
in this language,  at least for abelian states\cite{seidel3}.
Moreover, the efficiency of these 1D labels in deriving various
counting formulas for quasi-hole type states
has been suggested early on by
Haldane\cite{haldanetalk}, who arrived at the same
effective language by different means\cite{haldanetalk, bernevig}.
While such formulas can also be obtained by other
methods \cite{ardonne,read1},  Read has recently 
demonstrated agreement between the latter and the
1D approach\cite{read1}.
So far, however, the language of these 1D labels has been
discussed only for single component quantum Hall states.
Here we extend this language to the entire class of
Halperin $(m,m',n)$ bilayer states\cite{halperin}.
To this end, we develop a new technique that allows
the extraction of the general thin cylinder limits of these
states directly from the many-body wavefunctions.
Special attention is paid
to the class of $(m,m,m)$ states, and to the
$(331)$ state. For the latter, implications
concerning the possibility of a continuous phase
transition into a Moore-Read\cite{mooreread} (Pfaffian) state are discussed.
\\
\indent
{\bf Reduction to a 1D electrostatic lattice problem.}
In Refs. \onlinecite{seidel1, seidel2, karlhede2}, pseudo-potential Hamiltonians
have been used to determine the thin torus limits
of specific quantum Hall states. This method
proves to be unwieldy for a class as general
as the $(m,m',n)$ states, as it would require
high orders in perturbation theory for large
$m, m', n$. Here we derive a general method
to extract the limiting CDW patterns directly
from many-body wavefunctions.
Unless otherwise noted, we assume $m,m'>n$.
The $(m,m',n)$ states are expected to be $d\!=\!mm'\!-\!n^2$ fold degenerate
on the torus\cite{wen2}. Since 
the torus versions of the $(m,m',n)$
wavefunctions are quite complicated,
we work
on the cylinder, where the same CDW patterns must appear
between the edges
as the circumference approaches zero.
A straightforward generalization
of Halperin's $(m,m',n)$ state to cylindrical topology leads
to the expression
\begin{eqnarray}\label{mmn}
   \psi_{mm'n}=(\prod_I \xi_I^s)\, Q(\{\xi_i\};\{\xi_I\}) \times \exp(-\frac 12 \sum_\alpha {x_\alpha^2})
\end{eqnarray}
where $Q(\{\xi_i\};\{\xi_I\})
\!=\!\prod_{i<j} (\xi_i\!-\!\xi_j)^m \!\prod_{I<J} (\xi_I\!-\!\xi_J)^{m'}
 \!\prod_{i,J} (\xi_i\!-\!\xi_J)^n$. For $s\!=\!0$, the polynomial
part of \Eq{mmn} is thus the same that would appear in the usual disk
topology, with the particle coordinates
$z_\alpha\!=\!x_\alpha\!+\!iy_\alpha$ replaced
by functions $\xi_\alpha=\exp(\kappa z_\alpha)$ periodic in $y_\alpha$, where
$\kappa=2\pi/L_y$ and $L_y$ is the circumference of the cylinder.
We assume Landau gauge, $A=(0,x)$, for the vector potential,
letting the magnetic length equal to one. Uppercase (lowercase) indices
refer to particles in the upper (lower) layer, whereas Greek
indices refer to both layers. 
The integer $s$ in \Eq{mmn}
can be interpreted
as placing $s$ quasiholes in the upper layer at $x=-\infty$.
This does not affect the local properties of the incompressible
fluid described by \Eq{mmn}, which is confined to a finite
``ribbon'' on the infinite cylinder. It is left understood that
each such ribbon on the cylinder would be infinitely degenerate
by translation. We will see, however, that precisely $d$ distinct CDW patterns
are generated in the interior of the ribbon when the thin cylinder limit 
is taken, and when $s$ runs through
all possible values and translational symmetry is used.
\\
\indent
We will now compute the thin cylinder limit of
\Eq{mmn} using the following strategy.
The lowest Landau level (LLL) on the cylinder has a natural
basis of ring  shaped orbitals
$\psi_n= \xi^n\exp(-x^2/2-\kappa^2n^2/2)$  localized at a height
$x\sim \kappa n$ of the cylinder and
delocalized around the $y$ circumference.
We may expand the state \Eq{mmn} in
product wavefunctions $\psi_{n_1\dotsc n_N}=\psi_{n_1}(x_1)\cdot \dotsc
\cdot\psi_{n_N}(x_N)$ where
the $N$ particles occupy definite LLL orbitals.
As explained in Ref. \onlinecite{haldane1},
the coefficients in this expansion are of the
form
$\exp(\frac 12 \kappa^2 \sum_\alpha n_\alpha^2)
\cdot C(n_1\dotsc n_N)$, where $C(n_1\dots n_N)$
is the coefficient of the monomial $\xi_1^{n_1}\dotsc \xi_N^{n_N}$
in the polynomial part of \Eq{mmn}.
From this it is clear that those products
$\psi_{n_1\dotsc n_N}$ in the expansion of
\Eq{mmn} will dominate as $\kappa\rightarrow\infty$
for which the quantity
\begin{equation}\label{S}
  S=\sum_{\alpha=1}^N n_\alpha^2
\end{equation}
is maximal. Since the single particle
orbitals $\psi_n$ are well separated
along $x$ as $\kappa\rightarrow\infty$, the dominating
products $\psi_{n_1\dotsc n_N}$ describe a
well defined CDW pattern in the thin cylinder limit.
To identify this pattern, we observe that for any
monomial $\xi_1^{n_1}\dotsc \xi_N^{n_N}$ that has
a non-zero coefficient in the polynomial
$(\prod_i \xi_i^s)\, Q(\{\xi_i\};\{\xi_I\})$,
the exponents $n_\alpha$ satisfy the following
relation:
\begin{equation}\label{nalpha}
n_\alpha= \sum_\beta (m_{\alpha\beta} + p_{\alpha\beta}) + s\delta_{\alpha,\uparrow}
\end{equation}
where $m_{\alpha\beta}$ equals $m/2$ ($m'/2$) if both indices are in the
lower (upper) layer, and $n/2$ otherwise, $p_{\alpha\beta}$ is an {\em
antisymmetric} matrix whose values are restricted to
$\{-m_{\alpha\beta}, -m_{\alpha\beta}
\!+\!1,\dotsc,m_{\alpha\beta}\}$,
and we define $\delta_{\alpha,\uparrow}\equiv\sum_I \delta_{\alpha,I}$,
 $\delta_{\alpha,\downarrow}\equiv\sum_i \delta_{\alpha,i}$.
In \Eq{nalpha}, each choice for a particular
value of $p_{\alpha\beta}$ corresponds to the
choice of the term $\xi_\alpha^{m_{\alpha\beta}+p_{\alpha\beta}}\xi_\beta^{m_{\alpha\beta}-p_{\alpha\beta}}$ in the expansion of the factor
$(\xi_\alpha-\xi_{\beta})^{2m_{\alpha\beta}}$ in $Q(\{\xi_i\};\{\xi_I\})$.
We find that the choice of $p_{\alpha\beta}$ that maximizes
\Eq{S} is always of the form
\begin{equation}\label{pab}
  p_{\alpha\beta}=m_{\alpha\beta}\, \text{sgn}(\sigma_\alpha-\sigma_\beta)\;,
\end{equation}
where $\sigma$ is a permutation of $N$ objects.
A proof of this statement can be found in Ref. \onlinecite{v1}.
Figuratively speaking, \Eq{pab} says that each solution of
the maximization problem \Eq{S} corresponds to a ``ranking''
of the $N$ particles by means of the permutation $\sigma$.
If we imagine the $N$ particles arranged in a row according to their
ranking, we may call such an arrangement a state of 
``the squeezed lattice''. The term ``squeezed'' alludes to the
fact that, in contrast to the ``real space'' (orbital) arrangement
of the particles in the thin cylinder limit that we
are seeking, there are no empty sites in this squeezed
lattice. However, we will see below that the particle arrangements
on the squeezed lattice and in real space are closely 
related.
Via \Eq{pab}, the problem is reduced to finding a
permutation $\sigma$ 
that maximizes $S$.
 Since particles in the same
layer are identical, $S$ is invariant under
permutations of such identical particles on the squeezed lattice.
For our purposes, the state of the squeezed lattice
is thus fully described by the data 
$k_\alpha=\delta_{\bar\sigma_\alpha,\uparrow}-\delta_{\bar\sigma_\alpha,\downarrow}$,
which equal $+1$ or $-1$ depending on whether
the particle occupying the $\alpha$-th site,
$\bar\sigma_\alpha\equiv \sigma^{-1}(\alpha)$,
belongs to the upper or lower layer.
It turns out that in terms of the $k_\alpha$, $S$ can
be written as $S=-\frac 14 n(m+m'-2n) E$, where
\begin{equation}\label{E}
  E= \sum_{\alpha<\beta} -(\beta-\alpha) q_\alpha q_\beta
-2Q_0 \sum_\alpha \alpha q_\alpha + \text{const.}
\end{equation}
and $q_\alpha=k_\alpha(1+k_\alpha\Delta)$, $\Delta=(m'\!-\!m)/(m\!+\!m'\!-\!2n)$, $Q_0=(2s\!+\!N^\downarrow(n\!-\!m)\!+\!N^\uparrow(m'\!-\!n)\!+\!m\!-\!m')/(m\!+\!m'\!-\!2n)$, and we write $N^\downarrow$ ($N^\uparrow$) for the number
of particles in the lower (upper) layer.
\Eq{E} can be
interpreted as an electrostatic energy assigned
to each configuration of the squeezed lattice,
where a charge $q^{\uparrow,\downarrow}=\pm(1\pm\Delta)$
is assigned to upper and lower layer particles, respectively.
The first term in \Eq{E} is a linear 1D Coulomb
interaction between particles.
The second term can be interpreted
as a linear potential due to an external charge distribution,
e.g. a charge $+Q_0$ at the left boundary and a charge $-Q_0$
at the right boundary of the system.
Once the charge configuration minimizing $E$ is found,
the corresponding real space configuration,
i.e. the orbital position $n_\alpha$
of the $\alpha$-th
particle, follows from an ``un-squeezing'' rule
obtained from Eqs. \eqref{nalpha}, \eqref{pab}
\begin{equation}\label{unsqueeze}
n_\alpha=
\begin{cases}
m'N^{\uparrow}_\alpha+nN^\downarrow_{\alpha}+s & \mbox{if~~} k_\alpha=+1\\
mN^{\downarrow}_\alpha+nN^\uparrow_{\alpha} & \mbox{if~~} k_\alpha=-1\,,
\end{cases}
\end{equation}
where $N^{\uparrow}_\alpha$ ($N^{\downarrow}_\alpha$) is the number
of particles with $k_{\alpha'}=+1$ ($k_{\alpha'}=-1$) to the left of
site $\alpha$ on the squeezed lattice, i.e. for $\alpha'<\alpha$.
\Eq{E} can be viewed as a discrete version of the plasma
analogy in the thin cylinder limit. 
\\\indent
{\bf Solution of the electrostatic problem.}
We want to minimize the energy \Eq{E}  for given parameters
$m$, $m'$, $n$, and $s$.
It is useful to write $m-n=gr$, $m'-n=gr'$,
where $r$ and $r'$ are coprime integers. 
Since there are no vacancies on the squeezed lattice,
the particle configuration and thus the energy
$E$ is fully determined
by specifying the positions of the lower layer particles alone.
If $\alpha_j$ is the position of the
$j$-th lower layer particle on the squeezed lattice,
we find
\begin{align}
E&= q^\uparrow (q^\uparrow-q^\downarrow)\sum_{j=1}^{N^\downarrow}
(\alpha_j-x_j)^2 +\mbox{const,}\label{E2}\\
x_j&=\frac{r+r'}{r'}j-\frac{s}{gr'}-\frac{r}{r'}+\frac 12\;.\label{xj}
\end{align}
It is now clear how to minimize $E$ and thus maximize $S$.
We position the lower layer particles as closely
as possible to
the minima $x_j$, subject only to the constraint that
the positions $\alpha_j$ are integer, and that $1\leq
\alpha_1<\alpha_2<\dotsc<\alpha_{N^\downarrow}\leq N$.
This determines a pattern of lower and upper layer
particles in squeezed space. Certain boundary effects
may be present as follows. For example, the equilibrium
position of the first lower layer particle, $x_1$, may
be negative. Then for a range of $j$ values, the $x_j$'s
will be inaccessible by the respective particle
coordinates $\alpha_j$. This will lead to a clustering
of lower layer particles ($q^\downarrow$ charges)
at the left boundary of the squeezed lattice. This in turn
can be interpreted as a screening cloud to screen the
charge $+Q_0$ discussed below \Eq{E}.
However, for sufficiently large particle number
$j$ there will be a regime where the particle position
$\alpha_j$ can be placed within a distance $1/2$
or less
from $x_j$. This regime
of the squeezed lattice we will call the ''bulk''
as opposed to the screening cloud on either end of
the squeezed lattice. The appearance of different
regimes on the cylinder is expected for the general
wavefunction \Eq{mmn}, not only in the thin cylinder limit.
In particular, one may consider \Eq{mmn} with $N^\uparrow/N^\downarrow$
much different from the correct ratio $\nu^\uparrow/\nu^\downarrow$
of the upper and lower layer filling factors
in the $(m,m',n)$ phase, where $\nu^\uparrow=(m\!-\!n)/d$
and $\nu^\downarrow=(m'\!-\!n)/d$, respectively.
In this case \Eq{mmn} is known to describe a phase separated
state with phase boundaries between an $(m,m',n)$ bilayer phase
and a Laughlin-type phase in either the upper or lower layer \cite{yang}.
It is the bulk region of the squeezed lattice which corresponds
to the thin cylinder CDW pattern of the $(m,m',n)$
phase when the squeezed lattice is transformed
into real space via the un-squeezing rules \Eq{unsqueeze}.
We now derive some properties of these patterns.
First we note that some of the
$x_j$ may be half-odd integer. In this case both values
$\alpha_j=x_j\pm 1/2$ lead to a minimum of $E$.
The thin cylinder limit of the wavefunction \Eq{mmn}
is an equal amplitude superposition of the states generated
by all these configurations. In squeezed as well as in
real space, the patterns corresponding to the two cases
$\alpha_j=x_j\pm 1/2$ differ by an exchange between
two adjacent particles in different layers. For example,
for $m=5$, $m'=3$ and $n=2$ with $s=0$, 
we find the following
real space thin torus pattern from the procedure described
above:
\begin{equation}\label{531}
 \mbox{X}0\mbox{X}00\uparrow00\uparrow00\mbox{X}0\mbox{X}00\uparrow00\uparrow00\dots
\end{equation}
where characters denote the states of  consecutive
LLL orbitals, $0$ meaning unoccupied, and X$0$X represents
a triplet $\uparrow0\downarrow+\downarrow0\uparrow\;$.
The unit cell of the pattern \Eq{531}
has 11 sites. Using translational symmetry,
this fully accounts
for the ground state degeneracy of the $(532)$ state
on a torus.
The general situation is slightly more complicated.
From \Eq{xj}, we observe that when $j\rightarrow j+r'$,
$x_j$ increases by the integer $r+r'$, and
the pattern in squeezed space will repeat itself.
The squeezed space unit cell thus has size $r+r'$. It
consists of $r'$ charges $q_\downarrow$ and $r$ charges
$q_\uparrow$ and is thus neutral.
The size of the unit cell in real space can then be
obtained from \Eq{unsqueeze} with
$N^\uparrow_\alpha\rightarrow N^\uparrow_\alpha + r$,
$N^\downarrow_\alpha\rightarrow N^\downarrow_\alpha + r'$.
It is easy to see that both lines in \Eq{unsqueeze}
yield $n_\alpha\rightarrow n_\alpha+ (mm'-n^2)/g$.
The real space unit cell of the thin cylinder
pattern thus has size $d/g$.
Note that from this we correctly obtain $\nu^\uparrow=rg/d$,
$\nu^\downarrow=r'g/d$.
On a (thin) torus, a given pattern thus only accounts
for a fraction $1/g$ of the full degeneracy, using
translational symmetry. However, we see from \Eq{unsqueeze}
that varying $s$ will in general yield different
real space patterns.
By ``different'', we mean that the patterns are not
related by translation.
The patterns only repeat, up to
a translation,  when $s\rightarrow s+g$.
To see this, we note that one can find integers $a$, $b$
such that $a r = r' b+1$. If we let  $s\rightarrow s+g$,
$j\rightarrow j+a$ in \Eq{xj}, we find that $x_j$ is shifted
by the integer amount $a+b$. This means that the bulk
pattern on the squeezed lattice is merely shifted
by a constant. It is then easy to show from \Eq{unsqueeze}
that the real space bulk pattern is also just
shifted by a constant.
By varying $s$, we thus generate $g$ different bulk CDW
patterns in the thin cylinder limit, each having a
unit cell of size $d/g$.
On the torus these patterns, and those related by
translation, correspond to $g(d/g)=d$ distinct
ground states, as expected. We see that for
$g>1$, i.e. whenever $m'-n$ and $m-n$ are not coprime,
not all of the ground states on the torus are related
by translation. 
Furthermore, one may show from \Eq{xj} that spin fluctuations
as discussed for \Eq{531} are always present in precisely
one of the $g$ ground state patterns. 
In contrast,
 charge is always frozen
in the thin cylinder limit, and hence  
these fluctuations are unique to muti-component
systems.
In the following we discuss some
cases of special interest.\\\indent
{\bf $\boldsymbol {(m,m,m)}$ states. }
For $m=m'=n$ the quantity $S$,  Eq.
\eqref{S}, does not depend on the
permutation $\sigma$, \Eq{pab}. Every
configuration of upper and lower layer
particles on the squeezed lattice, for $N^\uparrow$, $N^\downarrow$
fixed, has the same amplitude in the
thin cylinder wavefunction. The
corresponding real space configurations
obtained from \Eq{unsqueeze} live
on a diluted lattice where every $m$-th
LLL-orbital is occupied.
The thin cylinder wavefunction is thus an equal amplitude
superposition of states on the diluted lattice
with given  $N^\uparrow$, $N^\downarrow$.
If the layer index is regarded as spin-$1/2$ index,
these thin cylinder wavefunctions are
ground states of the ferromagnetic Heisenberg Hamiltonian
$-J \sum_j (S_{mj} \cdot S_{(m+1)j}-1/4)$.
Using degenerate perturbation theory, we have
verified that the pseudo-potential
Hamiltonian whose exact ground state
is the $(m,m,m)$ state (see, e.g., Ref.\onlinecite{gima})
does indeed assume
this form in the thin cylinder limit,
to leading order in $\exp(-\kappa^2)$.
This shows that the low-energy sector
of the
$(m,m,m)$ pseudo-potential
Hamiltonian
has a gapless quadratically dispersing branch
even in this limit. By adding a generic
perturbation that breaks the $SU(2)$ symmetry down
to an easy-plane symmetry, the dispersion
of this mode becomes linear, as is well known
from the theory of spin-$1/2$ chains. Thus,
the low energy spectrum we obtain in the
thin cylinder limit completely agrees with
the spectrum expected for $(m,m,m)$ states
on infinite two-dimensional (2D) surfaces\cite{wenzee}.\\\indent
{\bf The $\boldsymbol{(331)}$ state.}
According to the general framework established
above, the eight $(331)$ ground states on the torus
come in two classes of four states.
These classes are distinct by the fact
that their respective members evolve into
one of the following two CDW patterns
in the thin
torus limit, up to
translations:\pagebreak
\begin{subequations}\label{331}
\begin{align}
&  \downarrow 0  \uparrow 0 \downarrow 0  \uparrow 0 \downarrow 0  \uparrow 0 \downarrow 0  \uparrow 0 \downarrow 0\uparrow 0\label{331_1}\\
&\mbox{X}\mbox{X}\; 0\, 0\; \mbox{X} \mbox{X}\; 0\, 0\; \mbox{X} \mbox{X}\;0\,0\,\mbox{X}\mbox{X}\;0\,0\;\mbox{X}\mbox{X}\;0\,0\,\label{331_2}
\end{align}
\end{subequations}
where again XX stands for $\uparrow\downarrow+\downarrow\uparrow$.
We have shown that indeed
both these states remain zero energy eigenstates
of the $(331)$ pseudo-potential Hamiltonian
(see, e.g., Ref. \onlinecite{gima})
in the thin torus limit when leading corrections
in $\exp(-\kappa^2)$ are considered.
As expected (c.f. Refs.\onlinecite{seidel1,seidel2,karlhede1,karlhede2}),
it is easy to show that one may form domain
walls between the two patterns in \Eq{331}
that carry the correct fractional charge
quantum numbers for Laughlin quasi-holes
in the $(331)$ state, e.g. $+1/8$ in the upper layer
and $-3/8$ in the lower. There is an intimate
connection between the wavefunctions of
the $(331)$ state and the $\nu=1/2$ Pfaffian state\cite{greiter}.
This is particularly suggestive
if we regard the thin torus CDW-patterns as the ''labels''
for the respective ground states, as discussed initially.
For, if we drop the spin index in the $(331)$-patterns \Eq{331}
by simply writing $1$ for each occupied site, we obtain $10101010\dotsc$
for \Eq{331_1} and $11001100\dotsc$ for \Eq{331_2}.
These are just the labels associated with the
$\nu=1/2$ Pfaffian\cite{karlhede2}.
Note that the latter is only sixfold degenerate on the torus,
as the four patterns related to \Eq{331_1} by translation
collapse into mere two patterns when spin indices are dropped.
The question of the possibility and nature of a continuous
(phase) transition between the $(331)$ state and the Pfaffian
has generated much interest\cite{ho,readrezayi}.
It is instructive to study this phase transition
in the thin torus limit. To this end, we examine
the effect of a small interlayer tunneling
term $H_t$ added to the pseudo-potential Hamiltonian
of the $(331)$ state.
We first consider the fate of the ground state \Eq{331_1}
when $H_t$ is added.
The  matrix elements of $H_t$ can be regarded as
spin flips. Writing $H=H_0+H_1+H_t$, where $H_0$
contains all terms of the pseudo-potential Hamiltonian
acting between identical or nearest-neighbor LLL-orbitals,
$H_1$ all remaining terms, we can treat $H_1+H_t$ in degenerate
perturbation theory in the thin torus limit.
Both states in \Eq{331} are ground states of
$H_0$, as well as all states related to \Eq{331_1} by
spin flips.
In the sector
spanned by states of the latter kind,
we find that to leading order in $\exp(-\kappa^2)$,
the effective Hamiltonian is given by
$ H_{eff} = J_z \sum_j \sum(\sigma^z_{2j}\sigma^z_{2j+2}+1)+t\sum_j \sigma^x_{2j}$,
where the $\sigma^{x,z}_{2j}$ are Pauli matrices, and $J_z\sim \exp(-2\kappa^2)$.
The hopping parameter $t$ is assumed to be much less than $\exp(-\kappa^2/2)$
(the gap of $H_0$)
but may be much larger than $J_z$.
$H_{eff}$ is a transverse field Ising model which has a well-known
phase transition at $t=J_z$. As a result of this 
transition,
the two ground states descended from \Eq{331_1} and its spin flipped
counterpart are replaced by a single ground state in the large $t$
phase. The same happens in the ground state sector related to
\Eq{331_1} by a single site 
translation.
 No transition
is found in the sector related to ground states of the type \Eq{331_2}.
It was already observed in Ref. \onlinecite{seidel2} that
ground states in different sectors will in general respond
differently to a 
perturbation of the pseudo-potential
Hamiltonian in the thin torus regime.
In this regime, the degeneracy of the two types of 
ground states shown in \Eq{331} is not maintained
as $t$ increases. 
Yet if we follow these ground states individually,
at large $t$ six of the eight ground states
evolve exactly 
into the six thin torus ground  states of the
$\nu\!=\!1/2$ Pfaffian, with spins polarized along the $x$ direction.
It is so far unclear
to what extent
the behavior
described here
can be extrapolated to the
regime of 2D quantum Hall phases. However, we argue
that any complete theory of a phase transition between
the $(331)$ state and the Pfaffian should reproduce this
behavior when the system is deformed into a thin torus.
In this sense, an intimate relation between this
``topological'' phase transition
and the transverse field Ising chain transition
may be suspected.\linebreak
\indent
{\bf Conclusion.}
We have derived a method to obtain the limiting
charge-density-wave form of Halperin $(m,m',n)$
bilayer
states on thin cylinders/tori. 
We found that basic properties of these
states such as filling factors, torus
degeneracies and fractional charges
can be obtained from a one-dimensional variant
of the plasma analogy.
In some important
special cases, a connection to the well known
physics of spin-$1/2$ chains has been made.
This also leads 
to a possible connection between phase
transitions in spin 
chains  
and in topologically ordered states of matter.


\begin{acknowledgments}
AS would like to thank Dung-Hai Lee for 
insightful discussions.
This work was supported by the state of Florida (AS), and NSF
grant No. DMR-0704133 (KY).
\end{acknowledgments}


\begin{thebibliography}{21}
\expandafter\ifx\csname natexlab\endcsname\relax\def\natexlab#1{#1}\fi
\expandafter\ifx\csname bibnamefont\endcsname\relax
  \def\bibnamefont#1{#1}\fi
\expandafter\ifx\csname bibfnamefont\endcsname\relax
  \def\bibfnamefont#1{#1}\fi
\expandafter\ifx\csname citenamefont\endcsname\relax
  \def\citenamefont#1{#1}\fi
\expandafter\ifx\csname url\endcsname\relax
  \def\url#1{\texttt{#1}}\fi
\expandafter\ifx\csname urlprefix\endcsname\relax\def\urlprefix{URL }\fi
\providecommand{\bibinfo}[2]{#2}
\providecommand{\eprint}[2][]{\url{#2}}

\bibitem[{\citenamefont{Laughlin}(1983)}]{laughlin}
\bibinfo{author}{\bibfnamefont{R.~B.} \bibnamefont{Laughlin}},
  \bibinfo{journal}{Phys. Rev. Lett.} \textbf{\bibinfo{volume}{50}},
  \bibinfo{pages}{1395} (\bibinfo{year}{1983}).

\bibitem[{\citenamefont{Seidel et~al.}(2005)\citenamefont{Seidel, Fu, Lee,
  Leinaas, and Moore}}]{seidel1}
\bibinfo{author}{\bibfnamefont{A.}~\bibnamefont{Seidel}},
  \bibinfo{author}{\bibfnamefont{H.}~\bibnamefont{Fu}},
  \bibinfo{author}{\bibfnamefont{D.-H.} \bibnamefont{Lee}},
  \bibinfo{author}{\bibfnamefont{J.~M.} \bibnamefont{Leinaas}},
  \bibnamefont{and} \bibinfo{author}{\bibfnamefont{J.~E.} \bibnamefont{Moore}},
  \bibinfo{journal}{Phys. Rev. Lett.} \textbf{\bibinfo{volume}{95}},
  \bibinfo{pages}{266405} (\bibinfo{year}{2005}).

\bibitem[{\citenamefont{Seidel and Lee}(2006)}]{seidel2}
\bibinfo{author}{\bibfnamefont{A.}~\bibnamefont{Seidel}} \bibnamefont{and}
  \bibinfo{author}{\bibfnamefont{D.-H.} \bibnamefont{Lee}},
  \bibinfo{journal}{Phys. Rev. Lett.} \textbf{\bibinfo{volume}{97}},
  \bibinfo{pages}{056804} (\bibinfo{year}{2006}).

\bibitem[{\citenamefont{Bergholtz and Karlhede}(2006)}]{karlhede1}
\bibinfo{author}{\bibfnamefont{E.~J.} \bibnamefont{Bergholtz}}
  \bibnamefont{and} \bibinfo{author}{\bibfnamefont{A.}~\bibnamefont{Karlhede}},
  \bibinfo{journal}{J. Stat. Mech.} \textbf{\bibinfo{volume}{L04001}}
  (\bibinfo{year}{2006}).

\bibitem[{\citenamefont{Bergholtz et~al.}(2006)\citenamefont{Bergholtz,
  Kailasvuori, Wikberg, Hansson, and Karlhede}}]{karlhede2}
\bibinfo{author}{\bibfnamefont{E.~J.} \bibnamefont{Bergholtz}},
  \bibinfo{author}{\bibfnamefont{J.}~\bibnamefont{Kailasvuori}},
  \bibinfo{author}{\bibfnamefont{E.}~\bibnamefont{Wikberg}},
  \bibinfo{author}{\bibfnamefont{T.~H.} \bibnamefont{Hansson}},
  \bibnamefont{and} \bibinfo{author}{\bibfnamefont{A.}~\bibnamefont{Karlhede}},
  \bibinfo{journal}{Phys. Rev. B} \textbf{\bibinfo{volume}{74}},
  \bibinfo{pages}{081308(R)} (\bibinfo{year}{2006}).

\bibitem[{\citenamefont{Rezayi and Haldane}(1994)}]{haldane1}
\bibinfo{author}{\bibfnamefont{E.~H.} \bibnamefont{Rezayi}} \bibnamefont{and}
  \bibinfo{author}{\bibfnamefont{F.~D.~M.} \bibnamefont{Haldane}},
  \bibinfo{journal}{Phys. Rev. B} \textbf{\bibinfo{volume}{50}},
  \bibinfo{pages}{17199} (\bibinfo{year}{1994}).

\bibitem[{\citenamefont{Seidel and Lee}(2007)}]{seidel3}
\bibinfo{author}{\bibfnamefont{A.}~\bibnamefont{Seidel}} \bibnamefont{and}
  \bibinfo{author}{\bibfnamefont{D.-H.} \bibnamefont{Lee}},
  \bibinfo{journal}{Phys. Rev. B} \textbf{\bibinfo{volume}{76}},
  \bibinfo{pages}{155101} (\bibinfo{year}{2007}).

\bibitem[{\citenamefont{Haldane}(2006)}]{haldanetalk}
\bibinfo{author}{\bibfnamefont{F.~D.~M.} \bibnamefont{Haldane}},
  \bibinfo{journal}{Bull. Am. Phys. Soc.} \textbf{\bibinfo{volume}{51}},
  \bibinfo{pages}{633} (\bibinfo{year}{2006}).

\bibitem[{\citenamefont{Bernevig and Haldane}(2007)}]{bernevig}
\bibinfo{author}{\bibfnamefont{B.~A.} \bibnamefont{Bernevig}} \bibnamefont{and}
  \bibinfo{author}{\bibfnamefont{F.~D.~M.} \bibnamefont{Haldane}},
  \bibinfo{journal}{cond-mat/0707.3637}  (\bibinfo{year}{2007}).

\bibitem[{\citenamefont{Read}(2006)}]{read1}
\bibinfo{author}{\bibfnamefont{N.}~\bibnamefont{Read}}, \bibinfo{journal}{Phys.
  Rev. B} \textbf{\bibinfo{volume}{73}}, \bibinfo{pages}{245334}
  (\bibinfo{year}{2006}).

\bibitem[{\citenamefont{Ardonne}(2002)}]{ardonne}
\bibinfo{author}{\bibfnamefont{E.}~\bibnamefont{Ardonne}}, \bibinfo{journal}{J.
  Phys. A} \textbf{\bibinfo{volume}{35}}, \bibinfo{pages}{447}
  (\bibinfo{year}{2002}).

\bibitem[{\citenamefont{Halperin}(1983)}]{halperin}
\bibinfo{author}{\bibfnamefont{B.~I.} \bibnamefont{Halperin}},
  \bibinfo{journal}{Helv. Phys. Acta.} \textbf{\bibinfo{volume}{56}},
  \bibinfo{pages}{75} (\bibinfo{year}{1983}).

\bibitem[{\citenamefont{Moore and Read}(1991)}]{mooreread}
\bibinfo{author}{\bibfnamefont{G.}~\bibnamefont{Moore}} \bibnamefont{and}
  \bibinfo{author}{\bibfnamefont{N.}~\bibnamefont{Read}},
  \bibinfo{journal}{Nucl. Phys. B} \textbf{\bibinfo{volume}{360}},
  \bibinfo{pages}{362} (\bibinfo{year}{1991}).

\bibitem[{\citenamefont{Keski-Vakkuri and Wen}(1993)}]{wen2}
\bibinfo{author}{\bibfnamefont{E.}~\bibnamefont{Keski-Vakkuri}}
  \bibnamefont{and} \bibinfo{author}{\bibfnamefont{X.-G.} \bibnamefont{Wen}},
  \bibinfo{journal}{Int. J. Mod. Phys. B} \textbf{\bibinfo{volume}{7}},
  \bibinfo{pages}{4227} (\bibinfo{year}{1993}).

\bibitem[{\citenamefont{Seidel and Yang}(2008)}]{v1}
\bibinfo{author}{\bibfnamefont{A.}~\bibnamefont{Seidel}} \bibnamefont{and}
  \bibinfo{author}{\bibfnamefont{K.}~\bibnamefont{Yang}},
  \bibinfo{journal}{arXiv:0801.2402v1}  (\bibinfo{year}{2008}).

\bibitem[{\citenamefont{Yang}(1998)}]{yang}
\bibinfo{author}{\bibfnamefont{K.}~\bibnamefont{Yang}}, \bibinfo{journal}{Phys.
  Rev. B} \textbf{\bibinfo{volume}{58}}, \bibinfo{pages}{R4246}
  (\bibinfo{year}{1998}).

\bibitem[{\citenamefont{Girvin and MacDonald}(1997)}]{gima}
\bibinfo{author}{\bibfnamefont{S.~M.} \bibnamefont{Girvin}} \bibnamefont{and}
  \bibinfo{author}{\bibfnamefont{A.~H.} \bibnamefont{MacDonald}},
  \emph{\bibinfo{title}{{Perspectives in Quantum Hall Effect, {\em edited by S.
  DasSarma, A. Pinczuk}}}} (\bibinfo{publisher}{Wiley, New York},
  \bibinfo{year}{1997}).

\bibitem[{\citenamefont{Wen and Zee}(1992)}]{wenzee}
\bibinfo{author}{\bibfnamefont{X.-G.} \bibnamefont{Wen}} \bibnamefont{and}
  \bibinfo{author}{\bibfnamefont{A.}~\bibnamefont{Zee}},
  \bibinfo{journal}{Phys. Rev. Lett.} \textbf{\bibinfo{volume}{69}},
  \bibinfo{pages}{1811} (\bibinfo{year}{1992}).

\bibitem[{\citenamefont{Greiter et~al.}(1992)\citenamefont{Greiter, Wen, and
  Wilczek}}]{greiter}
\bibinfo{author}{\bibfnamefont{M.}~\bibnamefont{Greiter}},
  \bibinfo{author}{\bibfnamefont{X.~G.} \bibnamefont{Wen}}, \bibnamefont{and}
  \bibinfo{author}{\bibfnamefont{F.}~\bibnamefont{Wilczek}},
  \bibinfo{journal}{Phys. Rev. B} \textbf{\bibinfo{volume}{46}},
  \bibinfo{pages}{9586} (\bibinfo{year}{1992}).

\bibitem[{\citenamefont{Ho}(1995)}]{ho}
\bibinfo{author}{\bibfnamefont{T.-L.} \bibnamefont{Ho}},
  \bibinfo{journal}{Phys. Rev. Lett.} \textbf{\bibinfo{volume}{75}},
  \bibinfo{pages}{1186} (\bibinfo{year}{1995}).

\bibitem[{\citenamefont{Read and Rezayi}(1996)}]{readrezayi}
\bibinfo{author}{\bibfnamefont{N.}~\bibnamefont{Read}} \bibnamefont{and}
  \bibinfo{author}{\bibfnamefont{E.}~\bibnamefont{Rezayi}},
  \bibinfo{journal}{Phys. Rev. B} \textbf{\bibinfo{volume}{54}},
  \bibinfo{pages}{16864} (\bibinfo{year}{1996}).

\end{thebibliography}

\end{document}